# Nonstationary photonic jet from dielectric microsphere

Yury E. Geints,[1,*] Ekaterina K. Panina,[1]
and Alexander A. Zemlyanov[1]

[1]V.E. Zuev Institute of Atmospheric Optics Siberian Branch of Russian Academy of Sciences, 1 Zuev square, Tomsk, 634021, Russian Federation
[*]ygeints@iao.ru

**Abstract:** A photonic jet commonly denotes the specific spatially localized region in the near-field forward scattering of a light wave at a dielectric micron-sized particle. We present the detailed calculations of the transient response of an airborne silica microsphere illuminated by a femtosecond laser pulse. The spatial area constituting the photonic jet is theoretically investigated and the temporal dynamics of jet dimensions as well as of jet peak intensity is analyzed. The role of morphology-dependent resonances in jet formation is highlighted. The evolution scenario of a nonstationary photonic jet generally consists of the non-resonant and resonant temporal phases. In every phase, the photonic jet can change its spatial form and intensity.

**1. Introduction**

A photonic jet (PJ) is a narrow high-intensity optical flux formed in the proximity of the shadow surface of transparent dielectric symmetric bodies (spheres, cylinders) with a diameter comparable to or somewhat larger than the wavelength of the incident optical radiation. The physical origin of PJ formation arises from the interference of the radiation net fluxes diffracted and passed through a particle. A most striking and specific feature of PJ is the extremely high spatial localization of the light field in the transverse direction (relative to the direction of incidence), which in contrast to the conventional high-NA focusing optics, can lead to sub-wavelength dimensions of the photonic jet. Thus, the PJ from micron-sized particle bears to the highest extent the diffractive features of wave-particle interaction process.

The investigations of PJ from symmetric particles were started from the theoretical work of Z. Chen et al. [1] where the authors for the first time noted that a jet-like light structure, which was formed in the geometric shadow of a transparent silica microcylinder exposed to a light wave experienced extremely high sensitivity to minor refractive index perturbations [2]. This specific field structure, called a *photonic jet* (or *photonic nano*jet [1]), is characterized by a sub-diffraction transverse dimension (smaller than a half-wavelength) and can extend conserving its shape to the distances of several wavelengths. This fact attracts a particular interest to the PJ phenomenon and benefits in wide practical applications, e.g., in designing the ultrahigh-resolution (nanometer-scale) optical sensors [3]. Besides, the PJ is used as surgical optical scalpel [4], optical tool for nano-object manipulation [5], element in optical data storage with ultrahigh resolution [6], and in the technology of nano-photolithography [7]. A rather thorough review of works devoted to theoretical and experimental investigations of the PJ phenomenon can be found in the recent paper [8].

As a rule, the classical scenario of PJ formation corresponds to the exposure of a dielectric microsphere to a continuous-wave radiation. This involves the use in numerical simulation the stationary Lorenz-Mie theory. At the same time, both optical technologies and biomedicine demonstrate recently the growing practical interest to laser systems generating ultra-short laser pulses. This type of laser radiation is attractive because it becomes possible to obtain extremely high peak intensity with low pulse energy. As a striking example, we can mention Ref. [9], where the authors have experimentally demonstrated the laser-induced perforation of living cell membranes using the photonic jet generated by micron-sized polystyrene spheres exposed to a femtosecond pulse of Ti:sapphire laser. It was noted that the proposed technology of femtosecond optical perforation, in contrast to the traditional technology using gold nanoparticles, did not lead to the death of a living cell.

At the same time, when the laser radiation is short the characteristic times of optical field formation inside and outside a scatterer can be comparable to the incident pulse duration. Thus, the transient scattering dynamics should be taken into account in the simulations. In previous works [10-12] based on internal optical field calculations was shown that the nonstationarity of light scattering at a transparent spherical particle manifests itself in temporal distortion of diffracted radiation and, in particular, in its temporal retarding. In the spectral domain, the scattering of an ultra-short and hence broadband optical signal on a dielectric particle may acquire the resonance character. The high-quality electromagnetic eigenmodes of a dielectric sphere, the so-called *whispering-gallery modes* (WGM), are excited in the internal optical field of a particle. As compared to the laser pulse duration, these modes have long lifetimes and they delay the incident radiation inside the particle [13] thus introducing the mentioned features into the light scattering.

In this paper, we theoretically investigate the temporal dynamics of the near-field scattering of ultra-short laser radiation at a transparent dielectric spherical microparticle, which results in the formation of nonstationary photonic jet. The numerical simulations based on the nonstationary Mie scattering theory are used to study the temporal evolution of the spatial shape and main characteristics of light jet.

**2. Theoretical model of nonstationary scattering of a laser pulse at a spherical particle**

Before considering the theoretical model and discussing the results of simulations, some remarks should be given. Below we will consider the linear scattering of a laser pulse at a micron-sized silica microsphere, which imposes certain constraints on the energy parameters of the incident radiation to avoid manifestation of optical nonlinearity of the particulate material. The strongest nonlinear optical effects in a solid dielectric such as silica are Kerr self-focusing and optical breakdown. In the near-IR wavelength region, the energy threshold of silica destruction by optical radiation is about 0.1-0.3 µJ, whereas the Kerr self-focusing power is at a level of ~ 2 MW [14]. Since analogous data for the visible region are unavailable in the literature, we will take these values as a basis for approximate estimation of nonlinear effect thresholds.

Then, for the present situation of irradiation of a particle with radius $a_0 = 2$ µm by a laser pulse with carrier wavelength $\lambda_0 = 0.4$ µm and temporal duration $t_p = 10$ fs we immediately obtain that the peak intensity should not exceed $I_0 \sim 10^{13}$ W/cm$^2$. Besides, if we take into account the focusing effect of optical radiation by the particle, this value should be additionally decreased by roughly three orders of magnitude. Thus, the condition of the absence of nonlinear effects in a silica particle during the laser pulse reads as $I_0 < 10^{10}$ W/cm$^2$.

In addition to the effect of optical nonlinearity upon ultra-short radiation scattering it is also important to take into account the chromatic dispersion of particle material, which manifests itself in the dependence of the linear refractive index $n$ on the laser wavelength $\lambda$. This leads to the changes in optical lens power of a particle proportional to the product $n(\lambda) a_0$ and to the spectral transformation (broadening/narrowing) of the pulse due to the group velocity dispersion during the propagation through the particle, which can be described by the coefficient $k''_\omega = \partial^2 k / \partial \omega^2 \big|_{\omega_0}$ ($\omega$ is laser circular frequency, and $k = 2\pi n / \lambda$ is the wave number in medium).

The chromatic dispersion of a material is usually described by the Sellmeier dispersion relation, and for fused silica it looks as follows [15]:

$$n^2(\lambda) = 1 + \sum_{m=1}^{3} A_m \frac{\lambda^2}{\lambda^2 - \lambda_m^2} \tag{1}$$

where $A_1 = 0.6962$, $A_2 = 0.4079$, $A_3 = 0.8975$, and $\lambda_1 = 0.0684$ µm, $\lambda_2 = 0.1162$ µm, $\lambda_3 = 9.8962$ µm. The laser pulse with the duration $t_p$ and carrier wavelength $\lambda_0$ has the spectrally limited half-width $\Delta\lambda = \lambda_0^2 / c t_p$, i.e., at $t_p = 10$ fs and $\lambda_0 = 0.4$ µm we obtain $\Delta\lambda = 0.06$ µm. In the wavelength range $\lambda_0 \pm \Delta\lambda$, Eq. (1) gives the corresponding variation of silica refractive index $n$ from 1.4650 to 1.4816 that can be neglected. In this situation, the second-order group velocity dispersion coefficient is $k''_\omega = 10^3$ fs$^2$/cm and consequently the silica dispersion in the 0.4-µm range is normal and leads to the twofold broadening of the pulse spectrum at the characteristic path length $L_D = t_p^2 / k''_\omega \sim 1$ mm, which exceeds the particle radius many time. Thus, in the following calculations we will ignore dispersion properties of silica particle.

To study the temporal evolution of optical pulse scattering at a particle we use the nonstationary Mie theory (NMT) developed by K. S. Shifrin et al. [10] being a combination of the Fourier spectral analysis and the stationary Lorenz-Mie theory [16]. In the framework of NMT, the initial nonstationary problem of the diffraction of a broadband radiation at a dielectric sphere reduces to the problem on the stationary light scattering of a set of monochromatic Fourier harmonics. The scattering properties of the particle are characterized by the so-called spectral response function $\mathbf{E}_s(\mathbf{r};\omega)$ being the traditional Mie series taken for all frequencies $\omega$ from initial pulse spectrum at every space point $\mathbf{r}$. The detailed description of this technique and some examples of its numerical realization can be found in Refs. [10-13]. Here, we briefly outline the basic points of the NMT and present the relations for time-dependent components of the optical field of scattered wave.

In numerical calculations, we use the following representation for the electric field of the linearly polarized along the $y$-axis incident plane optical wave:

$$\mathbf{E}^i(\mathbf{r};t) = \frac{1}{2}\left[\mathbf{E}^i(\mathbf{r};t) + \left(\mathbf{E}^i(\mathbf{r};t)\right)^*\right] = \frac{1}{2}E_0\mathbf{e}_y g(\tau)e^{j\omega_0\tau} + c.c., \quad (2)$$

where $g(\tau)$ is the temporal pulse profile, $\omega_0$ is the central frequency, $E_0$ is the real amplitude, $\tau = t - (z + a_0)n/c$ is the retarded time, $c$ is the speed of light in vacuum. The asterisk denotes the complex conjugate part. The dielectric spherical particle with the radius $a_0$ is located at the origin of coordinates, and the scattered laser pulse propagates in the positive $z$-direction. The temporal profile of the optical pulse is the Gaussian

$$g(\tau) = \exp\left\{-(\tau - t_0)^2 / 2t_p^2\right\}, \quad (3)$$

where $t_p$, $t_0$ are $e^{-1}$ pulse duration and initial time delay. Hereafter, for simplicity we use the complex representation of the fields with the complex conjugate part omitted.

To calculate the optical field distribution inside and outside the particle and to apply the results of the stationary Lorenz-Mie theory, it is necessary first to do transition from the temporal coordinates to the spectral domain using Fourier transformation of initial optical pulse

$$\mathbf{E}^i_\omega(\mathbf{r},\omega) = \Im[\mathbf{E}^i(\mathbf{r},t)] = \frac{1}{2}E_0\mathbf{e}_y G(\omega - \omega_0)e^{-jk_0(z+a_0)}. \quad (4)$$

Here, $\Im$ stands for the complex Fourier transformation, $G(\omega)$ represents the frequency spectrum of initial laser pulse, $k_0 = n\omega_0/c$.

Formally, Eq. (4) multiplied by the harmonic function $e^{j\omega t}$ determines the spectral component of the initial pulse in the form of a linearly polarized monochromatic wave with the amplitude

$$A(\omega) = E_0\mathbf{e}_y G(\omega - \omega_0). \quad (5)$$

The diffraction of this monochromatic wave at a spherical particle is described within the framework of the stationary Lorenz-Mie theory. The solution to this problem can be expressed as a Mie-series in spherical coordinates. Particularly, the expression for the external (scattered) electric field reads as

$$\mathbf{E}(\mathbf{r};\omega) = E_0 G(\omega - \omega_0) \sum_{l=1}^{\infty} R_l \left( j a_n(n, x_a) \mathbf{N}_{e1l}^{(3)}(\mathbf{r}) - b_n(n, x_a) \mathbf{M}_{o1l}^{(3)}(\mathbf{r}) \right), \quad (6)$$

where $R_l = j^l (2l+1)/(2l(l+1))$, and $\mathbf{M}_{o1l}^{(3)}$, $\mathbf{N}_{e1l}^{(3)}$ correspond to the vector spherical harmonics of the third kind. The series expansion coefficients $a_l$ and $b_l$ (the so-called Mie coefficients), depend on the relative refractive index (optical contrast) of the particle $n$ and the particle size-parameter $x_a = k a_0$:

$$a_l = \frac{\psi_l'(n x_a) \psi_l(x_a) - n \psi_l'(x_a) \psi_l(n x_a)}{\psi_l'(n x_a) \xi_l(x_a) - n \xi_l'(x_a) \psi_l(n x_a)}; \quad b_l = \frac{n \psi_l'(n x_a) \psi_l(x_a) - \psi_l'(x_a) \psi_l(n x_a)}{n \psi_l'(n x_a) \xi_l(x_a) - \xi_l'(x_a) \psi_l(n x_a)} \quad (7)$$

Here, $\psi_n(x)$ and $\xi_n(x)$ are the Riccati-Bessel functions, and the prime denotes the derivative with respect to the whole argument. The internal optical field can be written in the similar way [13].

Finally, the nonstationary electric field is expressed in the form of the convolution integral of initial pulse spectrum $G(\omega)$ and particle spectral response function $\mathbf{E}_\delta(\mathbf{r};\omega)$:

$$\mathbf{E}(\mathbf{r};\tau) = E_0 \left\{ G(\omega - \omega_0) \otimes \mathbf{E}_\delta(\mathbf{r};\omega) \right\}. \quad (8)$$

Here, $\mathbf{E}_\delta(\mathbf{r};\omega)$ denotes the Mie-series in the right-hand side of Eq. (6).

## 3. Discussion

Figures 1-3 show the results of our numerical calculations for the PJ dynamics formed in the vicinity of the airborne silica sphere with the radius $a_0 = 2.2$ µm exposed to a 10 fs laser pulse at carrier wavelength $\lambda_0 = 0.4$ µm. The power spectral response of the particle $S(\lambda) = |\mathbf{E}_\delta(\mathbf{r}_m, \lambda)|^2$ calculated at the point of the absolute intensity maximum of the internal optical field $\mathbf{r}_m$ at $\lambda_0$ is shown in Fig. 1. The three stacked plots in Figs. 2(*a*)-(*c*) demonstrate the temporal behavior of the main characteristics of the jet. The series of false-color images in Figs. 3(*a*)-(*l*) depicts the spatial 2D-distribution of the relative optical intensity $B(x, z) = |\mathbf{E}(x, z)|^2 / E_0^2$ within the PJ area at some selected time instants. The direction of radiation incidence is top-down. The factor $B$ in every 2D-distribution is normalized to its maximal value $B_{max}$ and is indicated in each image by the number.

*3.1 Peak PJ intensity*

Consider the temporal behavior of the peak PJ intensity in more detail. First of all, we address the spectral response of the particle near the laser carrier wavelength $\lambda_0$, which is shown in Fig. 1. The spectral profile of the initial pulse (in relative units) is also depicted in this figure for the reference. We can see that the function $S(\lambda)$ within the main part of the pulse spectrum changes insignificantly and consequently the temporal behavior of the intensity in the PJ-zone should be close to the temporal profile of the incident radiation. This is confirmed by the lower graph in Fig. 2, which shows the time history $B_{max}(t)$ of jet peak intensity together with the temporal pulse profile $I_0(t)$. As one can see, this similarity keeps true quite accurately till the time instant $t < 6t_p$ what is connected with the spatial remoteness of the PJ from the irradiated surface of the particle.

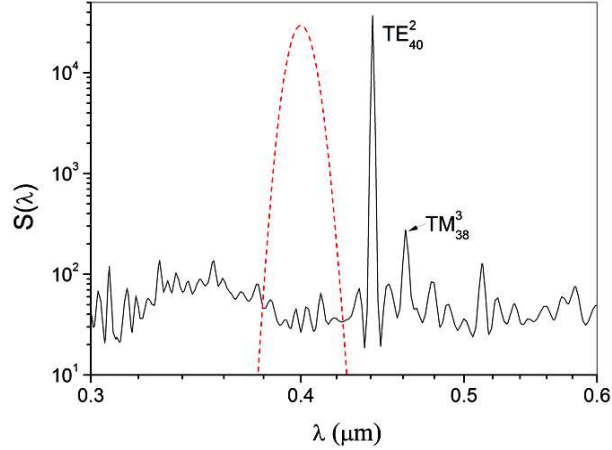

Fig. 1. Power spectral response of a silica sphere with $a_0$ =2.2 µm.
The initial pulse spectral contour is shown by dashed line.

At the same time, in the spectral response of the particle shown in Fig. 1 one can clearly see the excitation of the $TE_{40}^2$ morphology-dependent resonance (MDR) centered at $\lambda_r$ =0.4412 µm with a half-width of about 0.001 µm and $Q$-factor $Q \simeq 4800$. Mathematically, the resonances in the optical field of a dielectric sphere correspond to the situation when one of the denominators in Eq. (7) becomes minimal for a certain mode index $l$. This leads to a sharp increase in the absolute value of the corresponding coefficient, $a_l$ for TM-modes and $b_l$ for TE-resonances, as well as to the steeply increase of the internal field intensity. The mode order $l$ of the WGM indicated by the superscript shows the number of principal peaks in the radial distribution of the angle-averaged internal field intensity. Notice the tendency of $B_{max}$ values to decrease with an increase in the resonance order for a fixed mode number.

The MDRs are characterized by their quality factor $Q$ ($Q$-factor) showing how efficiently the optical energy is accumulated by the mode. The higher is the $Q$-factor of the resonance, the longer is the time period for which the eigenmode is capable to retain optical rays inside the resonator and the more strongly is the optical field confinement near the resonator walls. In transparent micron-sized spheres in the visible wavelength range, the characteristic values of (radiative) $Q$-factor usually have an order of magnitude as $Q \sim 10^4 \div 10^{10}$. As was noted above, the modes with the highest $Q$-factors are often called whispering-gallery modes.

Recalling the spectral response function, we note that according to our calculations the $TE_{40}^2$ resonance mode can capture only about 0.1% of the total pulse energy. Nevertheless, this WGM is excited and has a marked effect on the dynamics of optical intensity in the PJ area. Indeed, as seen from Fig. 2(*c*), the first and principal PJ intensity maximum caused by the laser pulse focusing at the particle is followed by pronounced intensity pulsations in time. These pulsations have exponentially decreasing amplitude and occur at later times when the incident laser pulse has already left the particle. In other words, after the primary PJ, the particle emits several less intense secondary photonic jets as a result of the pulse trapping in excited high-$Q$ WGMs. Since just the eigenmodes could be excited after the passage of the main pulse through the particle, the spectral parameters of secondary PJs are inherited by the spectral parameters of these eigenmodes, and in particular, they can be shifted in frequency with respect to $\lambda_0$.

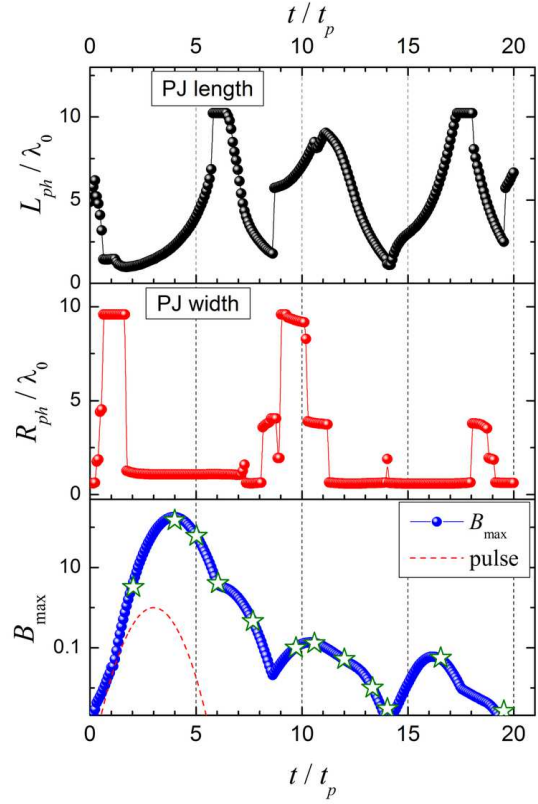

Fig. 2. Time-dependent PJ parameters; (a) length $L_{ph}$, (b) half-width $R_{ph}$, (c) peak intensity $B_{max}$. The star-symbols in (a) indicate the time instants for Fig. 3.

As can be seen, the temporal period of intensity pulsations is about 50 fs. Taking into account the fact that the excited WGM can be represented as a superposition of two counter propagating waves near the particle rim [12], the period of pulsations approximately corresponds to the time needed for optical wave to cover the particle circumference. Notice also the intensity beats, which are probably connected with the excitation of another resonance, namely $TM_{38}^3$ - mode having much lower $Q$-factor, which is indicated in Fig. 1 as well.

Thus, it should be emphasized that if a short-pulse laser radiation is scattered at a spherical particle, the intensity of the optical field in the PJ area may have be pulsed in time. The amplitude and the frequency of these pulsations depend on the parameters of particle MDRs excited by a laser pulse, as well as by the spectral detuning of these resonances from the central frequency of incident radiation.

*3.2 PJ shape and spatial dimensions*

In this section we will discuss the temporal dynamics of PJ shape and its spatial extent. For this purpose, we consider the 2D-distributions of the relative optical intensity near the shadow particle surface as plotted in Figs. 3(*a*)-(*m*). Every frame in this figure represents the spatial area with the dimensions 5.5×4 µm located in the equatorial section of the sphere in the plane of TE-mode polarization taken at a certain time instant marked by a "star" symbol in Fig. 2(*c*). For convenience, all images are grouped in three lines by four frames, and every line consists of intensity profiles calculated during the time period corresponding to the lifetime of a certain PJ intensity pulsation.

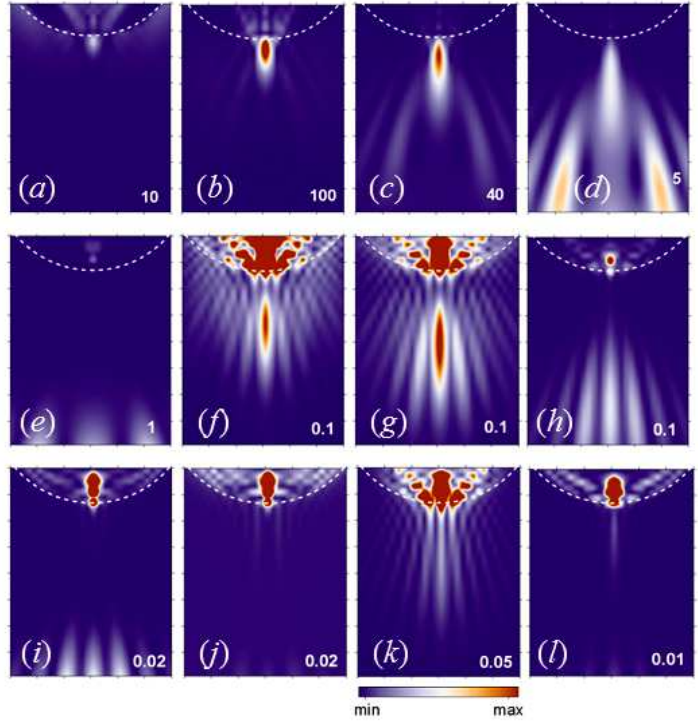

Fig. 3. Calculated 2D-distributions of normalized relative optical intensity $B/B_{max}$ in the PJ area at consecutive time instances (see text for details): $t/t_p$ = 2(a), 4(b), 5(c), 6(d), 7.7(e), 9.7(f), 10.6(g), 12(h), 13.3(i), 14(j), 16.5(k), and 19.5(l). Incident radiation is top-down; particle surface is indicated by dashed arc.

Initially (see Figs. 3(*a*)-(*c*)), the PJ of the classical shape is formed near the rear surface of the particle (shown by a dashed line), called in Ref. [17] the *dagger*-type jet without marked side lobes. The distinctive feature of this type of jets is that their intensity maximum is located right near the surface of the parent particle, and thus it seems that the PJ sticks to the particle.

The jet intensity begins to increase in time achieving the value $B_{max}$ =180 at $t \simeq 6t_p$, and simultaneously, the jet elongates up to its maximal length, $L_{ph} \approx 10\lambda_0$ (see Fig. 2(*a*)). At the same time, the PJ half-width, $R_{ph}$ measured at the $e^{-2}$ intensity level stably remains at the wavelength scale, $R_{ph} \simeq 0.8\lambda_0$. This spatial form and dimensions of the photonic flux resulting from the initial pulse focusing by a sphere correspond to the PJ formed in the stationary case [18].

However by the end of the initial pulse, i.e., roughly at $t > 6t_p$ (see Figs. 3(*d*),(*e*)) the photonic jet as a spatially localized structure practically disappears ($R_{ph} \gg \lambda_0$, $L_{ph} \sim \lambda_0$) and only several side lobes moving away from the particle remain visible.

The next temporal phase of PJ evolution comes with approximately 50 fs delay after the end of particle irradiation by a laser pulse (Figs. 3(*e*)-(*i*)). Here, the photonic jet is formed due to the field leakage of WGM excited off-resonance by outgoing laser pulse. The spatial shape of this secondary PJ changes cardinally, now looking like a bullet or a *flare* [17]. Visually, the photonic jet of this type is separated from the particle surface. The intensity of the secondary jet again first increases in time up to $B_{max} = 0.14$, i.e., to the level sevenfold lower than the peak intensity of the incident radiation, and then decreases vanishing in time for 30 fs (Fig. 3(*i*)). Along with the change in the morphological type, PJ's scales change as well. The PJ half-width nearly halves, going to the sub-wavelength region, $R_{ph} \simeq \lambda_0/2$ and PJ full length increases up to the value, $L_{ph} \simeq 20\lambda_0$ with allowance for the separation from the surface.

In addition, as can be seen from Figs. 3(*f*)-(*i*), the number of secondary PJ lobes increases significantly. In other words, the jet becomes *developed* in transverse direction. This circumstance indicates that the spherical particle no longer plays the role of an optical focuser to the incident radiation when the pronounced focal waist is formed and has the shape of outgoing light flow. Now, for light captured in particle eigenmodes the particle corresponds to an open optical cavity, i.e., the resonator with radiative energy losses. In accordance with the reciprocity principle these losses occur mostly through the shadow surface of the particle at the places of optical field antinodes (see Fig. 4), which results in the characteristic interference fringes in the PJ area.

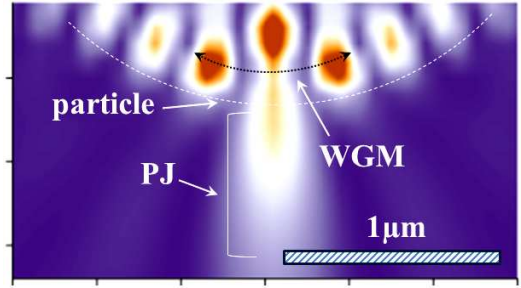

Fig. 4. Close-up of PJ area near the particle surface at $t/t_p = 24$.
The parameters of calculations are the same as in Fig. 3.

In this morphological type the photonic jet also exists at the next third stage of its evolution, when the captured in the WG mode optical wave after circumnavigating the particle rim again leaks out in the form of a light flow (lowest line of images in Fig. 3). It is clear that in this situation the PJ's length and width do not change but only PJ intensity decreases becoming approximately 15 times lower than the initial one, $B_{max} \simeq 0.06$.

The detailed calculations (not presented here) show that the following PJ evolution occurs similarly and is characterized by permanently decreasing intensity up to the complete resonance mode energy exhaustion. In the situation of $TE_{40}^2$-mode excitation by the wing of pulse spectral contour, its characteristic lifetime (inside the particle) is of the order of several picoseconds. However, just due to nonresonance character of mode excitation the intensity of the *external field* in the PJ zone decreases faster and approximately after 300 fs from the beginning of particle illumination the photonic jet "goes inside" the particle and becomes nearly indistinguishable. At this temporal stage, the evanescent field of the excited resonant mode in the structure of the near-field scattering prevails over the field of the outgoing wave [19] and the PJ terminates.

## 4. Conclusion

In conclusion, we have theoretically considered the scenario of nonstationary PJ development near the shadow surface upon scattering of a femtosecond laser pulse at a dielectric spherical micron-sized particle. Generally, this scenario consists of two temporal phases: nonresonance phase and resonance one.

The first nonresonance phase results from the forward scattering and focusing of the optical wave by the particle. The temporal behavior of the intensity in the PJ area generally resembles the temporal profile of the initial laser pulse. In the very peak of its development, the PJ has the spatial scales close to the characteristics of the conventional stationary photonic jet.

The resonance phase in PJ evolution comes later and is connected with the excitation and emission of electromagnetic eigenmodes in the optical field of a spherical particle. This temporal stage is characterized by the periodic pulsations of PJ intensity with permanently decreasing amplitude. The characteristic time of intensity decay is determined by the lifetime of the excited WGM having the highest quality factor. As compared to the first nonresonant temporal stage, the maximal PJ intensity during the resonant phase has lower values, but the jet width in the transverse direction is nearly halved.

The considered scenario of nonstationary PJ development is quite general, because it covers all possible situations of nonstationary elastic light scattering at a spherical particle. Actually, if a particle is rather small or a laser pulse is rather long so that no one high-Q resonance falls within pulse spectral profile, then the temporal PJ dynamics consists of only one first nonresonance phase. However, in the case when the incident radiation is scattered in resonance with some particle eigenmode, the dynamics of PJ formation includes both temporal stages but the phase with intensity pulsations may be sufficiently longer and the jet itself may be more spatially pronounced.

## Acknowledgements

We gratefully acknowledge funding from the Program No. 8.1 of the Division of Physical Sciences of Russian Academy of Sciences.